\documentclass[journal]{IEEEtran}
\usepackage{ragged2e,latexsym,amssymb,amsmath,graphicx,bbm}
\usepackage{epic}
\usepackage{psfrag}
\usepackage{amsbsy}
\usepackage[bbgreekl]{mathbbol}
\usepackage{epsfig}
\usepackage{cite}
\usepackage{setspace}
\usepackage{color}
\usepackage{ulem}

\newtheorem{lemma}{Lemma}[section]
\newtheorem{definition}{Definition}

\newtheorem{conjecture}{Conjecture}
\newtheorem{notation}{Notation}
\newtheorem{corollary}{Corollary}
\newtheorem{example}{Example}
\newtheorem{remark}{Remark}

\newtheorem{theorem}{Theorem}

\def\prob{\text{Prob}}
\renewenvironment{quotation}
               {\list{}{\listparindent=0pt
                        \itemindent    \listparindent
                        \leftmargin=10pt
                        \rightmargin=0pt
                        \topsep=0pt
                }
                \item\relax}
               {\endlist}
\begin{document}

\title{Optimized IR-HARQ Schemes Based on Punctured LDPC Codes over the BEC}
\author{
Iryna Andriyanova~\IEEEmembership{Member,~IEEE,}~and~
Emina Soljanin~\IEEEmembership{Senior Member,~IEEE}
}

\maketitle

\begin{abstract}
We study incremental redundancy hybrid ARQ (IR-HARQ) schemes based on punctured, finite-length, LDPC codes.
The transmission is assumed to take place over time varying binary erasure channels, such as mobile wireless
channels at the application layer. We analyze and optimize the throughput and delay performance of these
IR-HARQ protocols under iterative, message-passing decoding. We derive bounds on the performance that
are achievable by such schemes, and show that, with a simple extension, the iteratively decoded, punctured LDPC code based IR-HARQ protocol can be made rateless and operating close to the general theoretical optimum for a wide range of channel erasure rates.
\end{abstract}
\begin{IEEEkeywords}
HARQ, incremental redundancy, rateless codes, throughput vs.\ delay tradeoff, LDPC codes, puncturing, BEC.
\end{IEEEkeywords}

\section{Introduction}
In communications networks today, transmissions almost always take place over time varying channels,
because of, for example, the channel's physical nature (e.g., wireless) or the length of a session
(e.g., downloading a large file). Traditional channel coding schemes
are inadequate in such circumstances because they have fixed redundancy matching only a particular
channel condition. Similar problems arise in transmission to multiple users over (non)varying but
different channels. Several recently proposed and/or implemented coding schemes address the time varying and
multiuser communication scenarios, such as hybrid ARQ on the physical layer and Raptor codes
on the applications layer.

Hybrid ARQ transmission schemes  combine conventional ARQ with forward error correction.
A scheme known as {\it incremental redundancy} hybrid ARQ (IR-HARQ)
achieves higher throughput efficiency by adapting its error correcting code redundancy to varying channel conditions.
Because of that, the scheme has been adopted by a number of standards for mobile phone networks.
IR-HARQ is considered to be one of the most important features of the CDMA2000 1xEV-DO Release~0
and Revision~A systems~\cite{qiangwu02},~\cite{bushan06}.
A historic overview of HARQ schemes, up to 1998, can be found
in \cite{si98}. For a survey of more recent developments,
 we direct the reader to
\cite{itw07:coe} and references therein.
In the third generation wireless standards, the IR-HARQ scheme resides in the physical layer
and operates over time varying fading channels. The scheme is based on a turbo code dating back to
the IS-95 standard. A possible replacement of this code by an LDPC or a fountain code was considered
in \cite{SVW05:hybrid}.

Fountain codes are primarily designed to operate over erasure channels. They have superior performance
in applications in which the channel variations are large and/or cannot be reliably determined a priori.
Because of this robustness, some classes of Fountain codes have
been adopted into multiple standards, such as within the 3GPP MBMS standard for broadcast file delivery and streaming services, the DVB-H IPDC standard for delivering IP services over DVB networks, and DVB-IPTV for delivering commercial TV services over an IP network, and
are presently being considered for implementation in LTE eMBMS.

We here consider a hybrid ARQ scheme based on punctured LDPC codes over the BEC channel.
LDPC codes have been chosen as an instance of capacity-approaching codes.
They are theoretically well understood, and popular in practice
not only because of their error/erasure rate performance, but also because they have simple encoders and decoders.
In this particular application, capacity approaching LDPC codes are of interest because they can be punctured, as explained in \cite{4626154},
s.t.\ the resulting punctured ensemble is also capacity approaching.
Most of developed results can be easily extended to other punctured sparse-graph codes.

The performance of the HARQ scheme is measured by the throughput and the delay from the beginning of the coded data
transmission until the moment when the information has been successfully decoded. The goal is to have high throughput and low delay,
but only a certain tradeoff between these two quantities is attainable, and finding it is the central question in analyzing
HARQ schemes. One of our goals is to characterize the tradeoff between the average throughput and the average delay,
and to show how to run an HARQ scheme to achieve various operating points. Note that the average throughput and the average delay have been intensively investigated. However, the obtained results only give bounds, either under the maximum-likelihood decoding assumption (e.g. \cite{SVW05:hybrid, LZ08:harqII}), or under (more practical) iterative decoding but based on the bit error probability (e.g. \cite{SCV03:hybrid}),
which means that the bound is tight only for large code lengths. The approach taken in this paper is based on the {\it block} error performance under {\it iterative} decoding, and we use the finite-length scaling results on punctured LDPC code ensembles, as developed in \cite{AU09:punct-bec}. We also show how our LDPC codes based IR-HARQ scheme can be made rateless.

 The main contribution of this work are as follows: (i) We derive tight approximations of the average throughput and delay as functions of certain parameters of the used code ensemble and of the considered IR-HARQ scheme; (ii) We show how to chose these parameters in order to optimize both the throughput and the delay; (iii) We propose a rateless-like IR-HARQ scheme, based on LDPC codes, and derive tight bounds of its average throughput
 and delay.

This paper is organized as follows: In Sec.~\ref{sec:scheme}, we
describe our IR-HARQ scheme and present expressions for its average throughput and delay.
In Sec.~\ref{sec:code}, we define the finite-length rate-compatible LDPC codes used further in the paper.
Section \ref{sec:harq-ldpc} presents a model of the IR-HARQ scheme based on LDPC codes.
In Sec.~\ref{sec:optimization}, we define the optimization problem to determine the best code and protocol parameters.
Section \ref{sec:new} presents a modification of the IR-HARQ scheme based on LDPC codes, enlarging its working region, and the comparison of the modified scheme with the HARQ scheme, based on LT codes. At the end, in Sec~\ref{sec:discussion}, we then discuss our observations and some possible extensions.

\section{Incremental Redundancy Hybrid ARQ Model}
\label{sec:scheme}
\subsection{Multiple Transmissions Protocol and Channel Model}
\label{sec:mult-transm}
We analyze a particular retransmission protocol called {\it Incremental Redundancy Hybrid ARQ (IR-HARQ)}, with the
following multiple transmission model of \cite{emina:ruo:pred,AS09}: at the transmitter, the user data bits are encoded
by a low  rate code, referred to as the {\it mother} code. Initially, only a selected number of encoded bits are transmitted,
and decoding is attempted at the receiving end. If decoding fails, the transmitter, notified through the feedback, sends additional
encoded bits, thus incrementing the redundancy. Besides the information about the success/failure of the transmission, the feedback
may also carry the channel erasure rate information, to help the transmitter decide to which extent to increment the redundancy.
Upon completion of the new transmission, decoding is again attempted at the receiving end, where the
new bits are combined with those previously received.

The described procedure is repeated after each subsequent transmission request until all the encoded
bits of the mother code are transmitted. The channel is modeled as a time-varying BEC such that the channel erasure probability during the transmission of one block of encoded bits is constant and changes from one block transmission to another. We denote the channel erasure probability for transmission $m$ as $\epsilon_m$.
That the channel erasure probability does not change during the transmission of one block is a reasonable assumption as the block transmission duration is usually chosen to be smaller than the coherence time of the transmission channel.
This approach is used further in the paper, namely in Section \ref{sec:choice-param}, when the maximum number of transmissions is chosen.

The main design parameters of the IR-HARQ scheme are \cite{emina:ruo:pred,AS09}: the maximum possible number $M$ of transmissions for one block of user data and the fractions $q_m$, $m = \overline{1,M}$, of encoded bits assigned to transmission $m$. The maximum number of transmissions $M$ is usually predefined by the protocol, while the fractions $q_m$'s can be either predefined or calculated before each transmission, taking into account the feedback information about the previous channel erasure rates.

To analyze the IR-HARQ scheme, we adopt a probabilistic model in which the $q_m$'s are seen as probabilities, i.e., in which the transmitter assigns a bit to transmission $m$ with probability $q_{m}$.
Clearly, the transmitter has also the constraint (known as {\it rate compatible puncturing}) to assign
to transmission $m$ only those bits which have not been assigned to any of the previous transmissions.
Even with this probabilistic model it is possible to make the scheme rate compatible as follows \cite{emina:ruo:pred}:
\\
\\
{\tt START}\hfill\\
{\tt Before the IR HARQ protocol starts}
\begin{enumerate}{\leftmargin=3pt}
\item For each encoded bit, generate a number $\theta_v$ independently and uniformly at random over
$[0,1).$
\item Determine $M$ and $q_{1}$ (or all the $q_m$'s if necessary)
\item Compute $p_{1}$ as $p_{1} = 1 - q_{1}$.
\item Each bit s.t.\  $\theta_v \ge p_{1}$ is assigned to transmission $1$.
\end{enumerate}
{\tt If transmission $m-1$ fails for $2\le m<M-1$}
\begin{enumerate}
\item Determine $q_{m}$ (if not yet determined).
\item Compute $p_{m}$ as $p_{m} = p_{m-1} - q_{m}.$
\item Each bit s.t.\ $p_{m} \le \theta_v < p_{m-1}$ is assigned to
transmission $m$.
\end{enumerate}
{\tt If transmission $M-1$ fails}\hfill\\
\indent transmit all remaining bits.\\
{\tt END}\hfill
\vspace{1mm}

In the IR-HARQ transmission protocol above, the transmitter is assumed to have already accumulated some useful data to be sent, so the queuing process is not considered.

In Section \ref{sec:optimization} we determine how the $q_m$'s are chosen.
The criterion for such choice is to optimize the performance of the scheme, which is given by its throughput and delay.

\subsection{Performance Measures\label{sec:atd}}
Two standard measures of ARQ protocol efficiency are the {\it throughput} and the {\it delay}, defined as follows.
{\definition
The {\it throughput} of a retransmission scheme is the number of user data bits accepted at the
receiving end in the time required for transmission of a single bit.
}

{\definition
The {\it delay} of a retransmission scheme is the number of bits that must be transmitted in order to receive the useful
information (user data bits).
}

In what follows, we are interested by the {\it average throughput} $\eta$ and the {\it average delay} $\tau$.
We have the following lemma:

{\lemma
Consider an IR-HARQ scheme with at most $M$ transmissions and a set of fractions $q_1,\ldots,q_M$.
Let the underlying mother code be of length $n$ and of rate $R$.
Denote by $\omega_m$ the probability that it takes exactly
$m$ transmissions for the decoding to be successful.
Then the average throughput $\eta$ and delay $\tau$ are determined by following expressions
\begin{eqnarray}
\label{eq:eta}
\eta &=& \frac{R\displaystyle{\sum_{m=1}^{M}\omega_m}}{\displaystyle{\sum_{m=1}^{M} \omega_m
\Bigl(\sum_{j=1}^mq_j\Bigr)}};
\\
\label{eq:tau}
\tau &=& \frac{n \displaystyle{\sum_{m=1}^{M} \omega_m
\Bigl(\sum_{j=1}^mq_j\Bigr)}}{\displaystyle{\sum_{m=1}^{M}\omega_m}}.
\end{eqnarray}
}

\begin{proof}
The probability that one of the $m\le M$ transmissions is successful is $\sum_{m=1}^{M} \omega_m$. Because our protocol is limited to $M$ transmissions, if none of these transmissions is successful, the
throughput is equal to $0$. When one of the $m\le M$ transmissions is successful, the number of user data bits communicated to the receiver is $R n$.
The number of encoded bits sent to the receiver through the $m$th transmission is $n \sum_{j=1}^mq_j$.
So, the average throughput $\eta$ is given by (\ref{eq:eta}).
The calculation for $\tau$ is similar.
\end{proof}

{\remark
\label{remark1}
The expressions (\ref{eq:eta}) for $\eta$ and (\ref{eq:tau}) for $\tau$ implicitly assume that the feedback from the receiver to the transmitter is instantaneous. In practice the delay of the feedback transmission is positive,
and we can introduce it in the above expressions as follows.
Let the transmission time of one bit in the forward direction be $t_{1bit}$. Since the feedback propagation delay, i.e. the time interval between two transmissions, is $t$, it is equivalent to the time needed to transmit $n_{\text{\sc ACK}} = t/t_{1bit}$ bits in the forward direction.
Then the expression for $\tau$ becomes
\begin{eqnarray}
\label{eq:tau2}
\tau &=& n \frac{\sum_{m=1}^M \omega_m \sum_{ j} q_{j} +\frac{n_{\text{\sc ACK}}}{n} \sum_{m=1}^M m \omega_m}{\sum_{m=1}^M \omega_m},
\end{eqnarray}
where the term $\frac{n_{\text{\sc ACK}}}{n}\sum_{m=1}^M m \omega_m$ is proportional to the average feedback transmission delay.
This term grows with the number of transmissions.
On the other hand, note that the highest throughput can be achieved if the receiver is given a chance to attempt decoding upon receiving each additional bit, that is when $M=n$.
}

The expression (\ref{eq:tau}) for throughput becomes equal to its counterpart in \cite{SCV03:hybrid} when $q_m=1/M$.
The authors of \cite{SCV03:hybrid} expressed the quantity $\omega_m$ in terms of the probability $P(m)$ that the asymptotic\footnote{i.e., when the codelength $n \rightarrow \infty$.}
bit erasure rate $P_b$ at transmission $m$ goes to $0$, i.e., $P(m)\approx \text{Prob}[ P_b^{(m)} \rightarrow 0].$
For LDPC codes, this probability has been computed with the help of density evolution.
Clearly, $P(m)$ is a lower bound on the failure probability at transmission $m$, which thus gives an upper bound on $\eta$ and a lower bound on $\tau$. We next derive expressions for these asymptotic bounds, while tighter bounds for finite length case will be presented in Section \ref{sec:harq-ldpc}.

\begin{figure}[hbt]
\begin{picture}(0, 400)
\put(30,185){\includegraphics[scale=0.9]{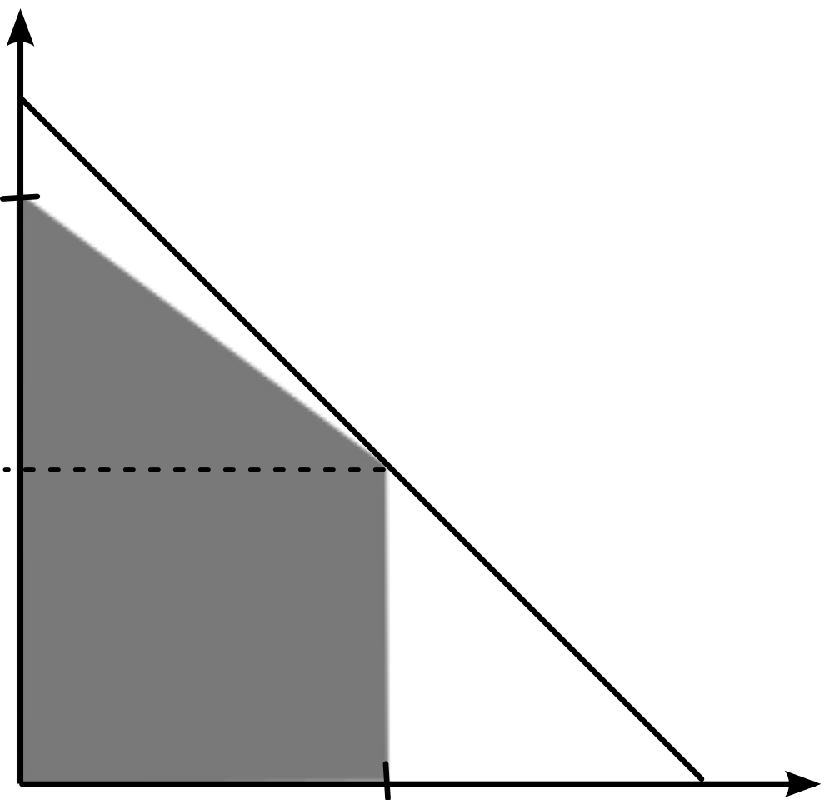}}
\put(30,0){\includegraphics[scale=0.34]{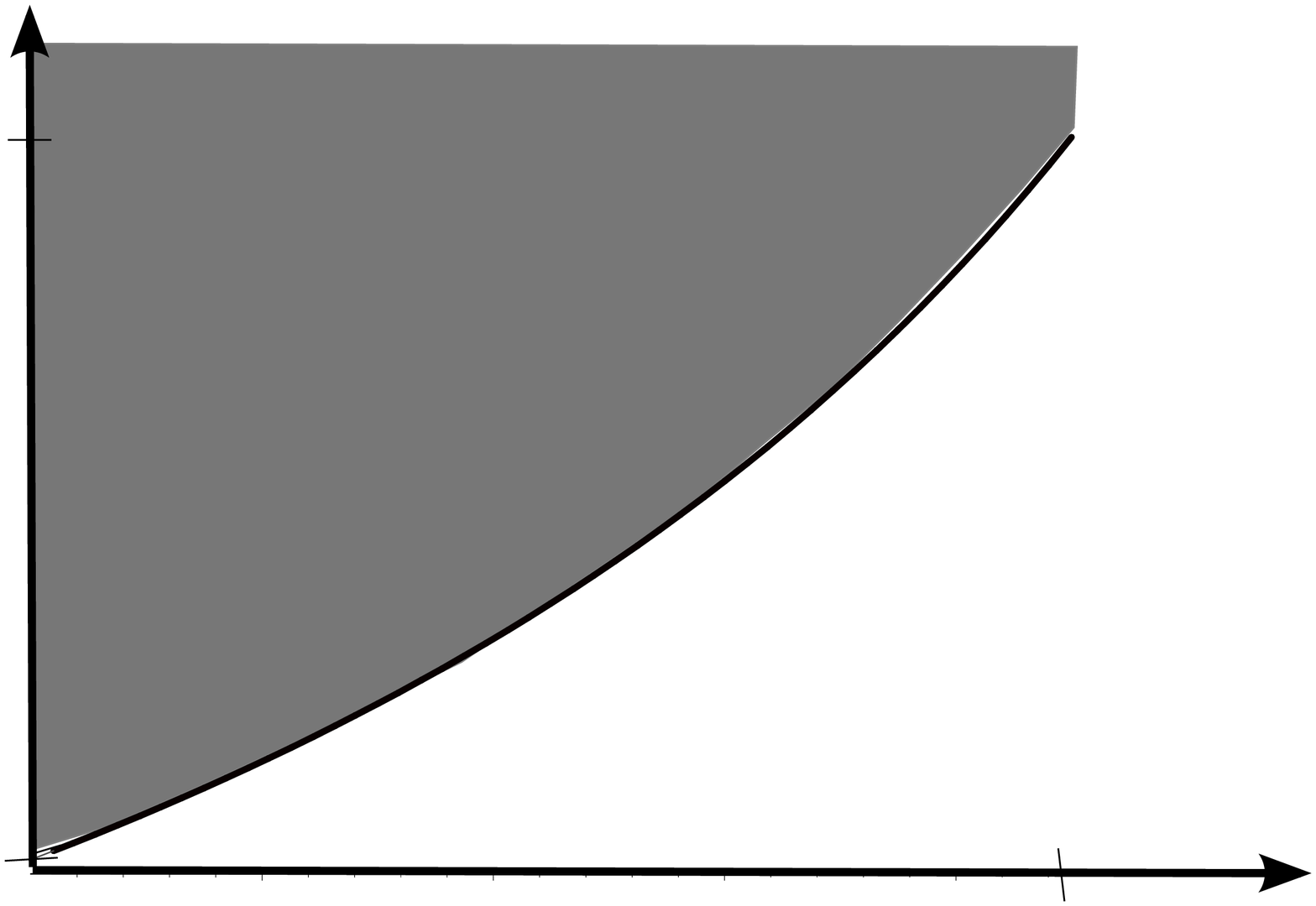}}
\put(22,385){$\eta$} \put(240,180){$\epsilon$}
\put(25,365){$1$} \put(210,180){$1$} \put(115,300){$\eta_{max}=1-\epsilon$}
\put(6,340){$\frac{R}{1-\epsilon^*_{(n)}}$} \put(21,269){$R$} \put(120,180){$\epsilon^*_{(n)}$}
\put(20,8){$\tau_0$} \put(22,128){$n$} \put(22,150){$\tau$}
\put(208,-5){$\epsilon^*_{(n)}$} \put(240,-5){$\epsilon$}
\end{picture}
\caption{ \label{fig:upbound} Illustration of the upper bound on the region of attainable throughputs (top) and of the lower bound on the region of attainable delays (bottom) for an IR-HARQ scheme over a BEC($\epsilon$), as shown in Thm.~\ref{thm:bounds}. The scheme is based on a length-$n$ LDPC code of rate $R$ and iterative threshold $\epsilon^*_{(n)}$. The channels capacity line $\eta_{max}=1-\epsilon$ is the maximum attainable throughput. The delay at $\epsilon=0$ is $\tau_0=n(1-\epsilon^*_{(n)})$.}
\end{figure}

Consider an example of sparse-graph codes. A randomly chosen code from a sparse ensemble of length $n$ has a successful iterative decoding with high probability when the channel erasure probability $\epsilon$ is smaller than  $\epsilon^*_{(n)}$, where $\epsilon^*_{(n)}$ is the so called finite-length iterative decoding threshold. We will discuss $\epsilon^*_{(n)}$ for a particular case of LDPC codes in Section \ref{sec:code}.
Now we can state the following result:
{\theorem \label{thm:bounds}
Consider an IR-HARQ scheme based on a sparse-graph code of rate $R$ and
iterative decoding threshold $\epsilon^*_{(n)}$.
The following bounds hold:
\begin{eqnarray}
\label{eq:UB-eta}
\eta &\leq& \begin{cases}
R\frac{1-\epsilon}{1-\epsilon^*_{(n)}},& 0\leq \epsilon \leq \epsilon^*_{(n)};\\ 0,& \text{otherwise};
\end{cases}
\\
\label{eq:LB-tau}
\tau &\geq&  \begin{cases}
n \frac{1-\epsilon^*_{(n)}}{1-\epsilon},& 0\leq \epsilon \leq \epsilon^*_{(n)};\\ \infty,& \text{otherwise}.
\end{cases}
\end{eqnarray}
}
\begin{proof}
Consider the limiting case $M=n$ (that is bit-by-bit transmission) since the highest throughput can be achieved if the receiver is given
a chance to attempt decoding upon receiving each additional bit, that is when $M=n$.
The smallest fraction of bits that are sufficient for successful decoding is $1-\epsilon^*_{(n)}$.
The channel with erasure probability $\epsilon<\epsilon^*_{(n)}$ passes on average a fraction of $1-\epsilon$ bits unerased.
Hence, the smallest fraction $\gamma$ of coded bits to be sent by the transmitter in order to receive a fraction of $1-\epsilon^*_{(n)}$
bits on average is
$$\gamma = \frac{1-\epsilon^*_{(n)}}{1-\epsilon}.$$
Note that
$\eta \leq R/\gamma$, and (\ref{eq:UB-eta}) follows immediately.

Now consider the case when $M=1$ and $n \rightarrow \infty$.
At least $\gamma n$ bits should be sent to ensure successful decoding.
Hence, $\tau \geq \gamma n$ and (\ref{eq:LB-tau}) follows.
\end{proof}
The derived bounds are illustrated for an example LDPC ensemble in Fig.~\ref{fig:upbound}.

\section{Performance of (Punctured) Finite-Length LDPC Codes over the BEC}
\label{sec:code}

As we have seen above, the performance of the IR-HARQ scheme depends on the decoding performance after each transmission.
We assume that the mother code is an LDPC code.
We will see later that the performance after each transmission in this case is related to the decoding performance of the
punctured mother code. First let us define the mother code and describe the puncturing technique.

\subsection{The Mother Code and Puncturing}
The mother code is taken at random from an irregular length-$n$, LDPC code ensemble, defined by its degree distributions $\lambda(x)=\sum_{i}\lambda_ix^i$ and $\rho(x)=\sum_{j}\rho_jx^j$.\footnote{We refer the reader unfamiliar with LDPC codes and their properties that we use below to the textbook \cite{RicUrb:book}.}
Each code in the ensemble corresponds to a different Tanner graph, having
$\lambda_i$ fraction of edges incident to variable nodes of degree $i$ 
and $\rho_j$ fraction of edges incident to check nodes of degree $j$ respectively. 

A code taken at random from an ensemble of $(\lambda,\rho)$-LDPC codes has, with high probability, a bit error probability close to the average bit error probability $P_b$ of the ensemble. We will refer to this property as concentration.
This property implies the concentration of the block error probability $P_B$ for the so called waterfall region of channel parameters, 
within which $P_B \propto P_b$.
The concentration property allows us to only consider the average performance of an LDPC ensemble (instead of looking at the performance of a particular code) by using the ensemble average analysis techniques.

The performance of iterative decoding averaged over the LDPC ensembles is well understood when $n$ is sufficiently large and when LDPC codes are used for a transmission over a channel
with some fixed erasure rate $\epsilon$. Namely, as long as the channel erasure rate $\epsilon$ is smaller than the threshold value
$\epsilon^{\ast}$ given by
\[
\epsilon^{\ast} = \min_{x\in(0,1]}\frac{x}{\lambda(1 - \rho(1 - x))},
\]
the iterative message passing algorithm leads to vanishing bit-erasure probability as the number of iterations grows.

Puncturing is a technique to obtain a code of a higher rate from a given code of some rate $R$. It simply means not transmitting (puncturing) a fraction of the encoded bits. The performance of the resulting code depends on the number and the choice of punctured bits.
One way to make this choice is at random, depending on the outcome of tossing the same (biased) coin for each variable node.
This way of puncturing is often called random puncturing.

Another way to select the bits to puncture is to first choose the degree of the node to be punctured, according to a certain (optimized, degree biased) probability distribution, and then to select a node to puncture uniformly at random from all nodes with the chosen degree. This way of puncturing is often referred to as intentional puncturing.
It has been shown \cite{HKM04:rate-compatible} that intentional puncturing outperforms random puncturing, and,
even more importantly, it can be designed to conserve the concentration property, whereas random puncturing cannot.
Therefore, in what follows we only consider intentional puncturing.

A {\it punctured} LDPC ensemble of some length $n$ is described by three polynomials: the degree distributions $\lambda(x)$ and $\rho(x)$ mentioned before and the puncturing degree distribution
$p(x)=\sum_i p_i x^{i-1}$, where the $p_i$'s are the probabilities with which variable nodes of degree $i$ are punctured.
{\notation \label{notation1} Let
${ \lambda_p(x)} = \sum_i p_i \lambda_i x^{i-1}$ and \\
${ \bar \lambda_p(x)} = \sum_i (1-p_i) \lambda_i x^{i-1}=\lambda(x)-\lambda_p(x)$.
}
Using this notation, the asymptotic iterative threshold of such a punctured LDPC ensemble that was obtained in \cite{AU09:punct-bec},
becomes
\begin{equation} \label{eq:de}
\epsilon^{\ast} = \min_{x\in(0,1]}\frac{x - \lambda_p(1 - \rho(1 - x))}{\bar \lambda_p(1 - \rho(1 - x))},
\end{equation}
and its design rate is given by
\begin{equation} \label{eq:design-rate}
R_p = \frac{R}{1-\frac{\sum_{i} p_i \lambda_i /i}{\sum_{i}\lambda_i/i}},
\end{equation}
where $R$ is the code rate of the mother ensemble.

\subsection{Finite-Length Performance}
\label{app:param}
We start with introducing some useful notation which we need to present finite-length performance of (punctured) LDPC codes.
{\notation
Note that the fraction of the variable nodes of degree $i$ of
 a $(\lambda,\rho)$ LDPC ensemble is $\Lambda_i=i \lambda_i/ (\sum_i i \lambda_i)$, $1\le i\le k$
We denote by ${ \Lambda(x)} =\sum_i \Lambda_i x^{i}$ the variable node degree distribution.
Also, given the puncturing degree distribution $p(x)$, let
${ \bar \Lambda_p(x)} = \sum_i (1-p_i) \Lambda_i x^{i-1}.$
}

{\notation
Finally, we introduce the following notation:
\begin{eqnarray*} { y(x)}=1-\rho(\bar x),&
{ \pi (y)}  = \epsilon \bar \lambda_{p} (y)+\lambda_{p} (y),\\
{ \xi(x)} = (\pi'(y))^2 (\bar y)(\rho'(1)-\rho'(\bar x)),&
{ \mu(x)}=\pi'(y)\rho'( \bar x),
\end{eqnarray*}
where $\bar x = 1-x$ and $\bar y=1-y$.
Here and further in the paper primes denote derivatives.
}

The following conjecture from \cite{AU09:punct-bec} will be further used:
\begin{conjecture}
\label{conj:1}
Assume transmission takes place over the BEC with erasure probability $\epsilon$ using a code chosen at random from a punctured LDPC ensemble with length $n$ and puncturing degree distribution $p(x)$.
Then, with high probability, the block erasure rate is tightly approximated by the following expression
\begin{equation} P_B =  Q\left( \frac{\sqrt{n}(\epsilon^*-\epsilon-\beta n^{-2/3})}{\alpha} \right) + o(1),
\label{eq:app}
\end{equation}
where $Q(\cdot)$ is the Q-error function and
$\alpha$ and $\beta$ are the scaling and shift parameters, given by
\begin{eqnarray}
\label{eq:alpha}
\alpha &=& \sqrt{
\frac{\xi (x^*)}{ \Lambda'(1) }}  \left(
\frac{1}{\bar \lambda_{p}(y^*)}-\right. \nonumber \\ &&\left.
\frac{2 \bar \lambda'_{p} (y^*) \rho'(1- x^*)(1-\mu(x^*))}{\bar \lambda_p(y^*)^2  \cdot \mu'(x^*)}
\right),
\\
\label{eq:beta}
\beta&=& {\left( \frac{ {b}}{  \bar \Lambda'_p(y^*)  x^* \rho'(1- x^*) \cdot \sqrt{ -  \bar \lambda_p(y^*)  \mu'(x^*) }  } \right)}^{2/3},
\end{eqnarray}
where
$x^*$ satisfies (\ref{eq:de}), $y^*=y(x^*)$, and
\begin{multline*}b=
x^* \rho'(1- x^*)
 \frac{\lambda'(y^*)}{\lambda(y^*)}
\frac{y^*-x^* \rho'(1- x^*)}{y^*}\\
+ \frac{\left(x^* \rho'(1- x^*)\right)^2}{\pi(y^*)} \left(  \pi''(y^*)+\frac{\pi'(y^*)}{y^*}-\frac{ \pi(y^*)'^2}{\pi(y^*)}\right)\\
+ \left(\frac{x^* (1-\epsilon^*) \rho'(\bar x^*)}{y^*}\right)^2 \cdot
 \frac{\sum_l l p_{l} (1-p_{l}) \lambda_{l+1}y^{*(l-1)}}{\pi(y^*)}.
\end{multline*}
\end{conjecture}

As we can see, $\alpha$ and $\beta$
 only depend on $\epsilon^*$, $x^*$ and $y^*$, as well as on polynomials $\lambda$, $\rho$ and $p$.
 The justification for the conjecture follows the same line of reasoning as for the conjecture of the finite-length scaling law for unpunctured LDPC codes in \cite{4777618}. Note that the conjecture for unpunctured LDPC codes has been proven in \cite{dembo-mont} for a particular case.

{ \example [Regular codes]
\label{example:reg}
For regular LDPC codes with parameters $\lambda(x)=x^{c}$ and $\rho(x)=x^d$,  we have that $p(x)=p x^c$, where $0\leq p \leq 1$. Moreover,
the performance parameters become \begin{equation}
\label{eq:ab-reg} \epsilon^*= \frac{\epsilon^*_0}{1-p}, \qquad \alpha = \frac{\alpha_0}{1-p},\qquad
 \beta = \frac{\beta_0}{1-p},
\end{equation}
where $\epsilon^*_0$, $\alpha_0$ and $\beta_0$ are the parameters of the corresponding unpunctured ensemble.
}

{\remark
\label{remark:1}
For an LDPC code ensemble of length $n$, the finite-length iterative threshold $\epsilon^*_{(n)}$, already mentioned in Section \ref{sec:scheme}, is \cite{4595227}
$$\epsilon^*_{(n)} = \epsilon^*-\beta n^{-2/3}.$$
Note that, even for moderate lengths $n$, $\epsilon^*_{(n)}$ lies close to the asymptotic threshold $\epsilon^*$.
}
\subsection{Equivalent Puncturing Model of the IR-HARQ Scheme Based on LDPC Codes}
\label{sec:harq-ldpc}
Consider the IR-HARQ scheme described in Section \ref{sec:scheme}.
Its mother code is an LDPC code chosen at random from the ensemble of given length $n$, with degree distributions $\lambda(x)$ and $\rho(x)$.
Since it is irregular, the IR-HARQ scheme is now parametrized by the maximum number of transmissions $M$ and the sequence of $q_{ij}$'s and $j=\overline{1, M}$, where $q_{ij}$ denotes the probability with which a bit of degree $i$ is chosen for transmission $j$.

Recall that in Section \ref{sec:scheme}, only one value $q_j$ was assigned to transmission $j$.
However, if the bits of an irregular code were chosen to be transmitted with probability $q_j$ regardless of their degree, this would correspond to random puncturing and the concentration property would be lost \cite{AU09:punct-bec}.
By introducing $q_{i j}$, for variable nodes of degee $i$ and transmission $j$, we obtain the intentional puncturing scheme and preserve the concentration of the code performance around the average.

From now on, we consider the puncturing model based on the $q_{ij}$'s. Note that, for this case, the term $\sum_j q_j$ in expressions (\ref{eq:eta}), (\ref{eq:tau}) and (\ref{eq:tau2}) should be replaced by $\sum_{i,j} \lambda_i q_{i,j}$.

The IR-HARQ protocol can be described with the help of the following equivalent punctured code model:
the bits that the transmitter chooses to send through the $m$-th transmission
can be equivalently seen as obtained by implementing a puncturing device that punctures a bit corresponding
to a variable node of degree $i$ with probability $p_{im}$, where $p_{im}=1- \sum_{j = 1}^{m}q_{ij}$, or, as shown
within the protocol described in Section \ref{sec:mult-transm},
\begin{equation}
\label{eq:pq}
p_{i1} = 1 - q_{i1} ~ \text{and} ~
p_{ij} = p_{i(j-1)} - q_{ij} ~\text{for} ~ 1< j \le M.
\end{equation}
Further, assume that a transmission $j$ takes place over the BEC with probability $\epsilon_j$.
When a bit corresponding to a variable node of degree $i$ is assigned to one of the first $m$ transmissions,
it can be viewed as passing through the channel with average erasure rate\footnote{In this case, it is assigned to transmission $j$ with probability $q_{i j}/(\sum_{k=1}^m q_{ik})$.}
$(\sum_{j=1}^m q_{i j} \epsilon_j)/(\sum_{k=1}^m q_{ik})$. So we can model the IR-HARQ
protocol through transmission $m$ as the transmission of the punctured mother code over a BEC with average erasure rate
\begin{equation} \label{eq:eps-m}
\delta_m=\sum_i \lambda_i \frac{\sum_{j=1}^m q_{i j} \epsilon_j }{\sum_{k=1}^m q_{ik}},
\end{equation}
where the considered bit is punctured with probability $p_{im}$.

The IR-HARQ protocol outlined below implements our model while conforming to the rate compatible puncturing; it is based on the one introduced in Section \ref{sec:mult-transm}.

Since the $q_{im}$'s are linked to the $p_{im}$'s, the IR-HARQ performance can be determined from the performance of punctured versions  of the mother code.
We now determine the expected throughput and delay of the IR-HARQ scheme.
Consider expressions (\ref{eq:eta}) and (\ref{eq:tau}).
First we switch to the irregular case by replacing $q_m$ by $\sum_i\lambda_i q_{im}$.
Next we describe how the $\omega_m$'s can be determined.

Let $A_m$ denote the event of successful decoding after $m$ transmissions, so $\bar A_m$ denotes a decoding failure.
Then
\begin{eqnarray*}\omega_m&=&\prob(\bar A_{m-1})\prob(A_m|\bar A_{m-1})\\
&=&\prob(\bar A_{m-1})-\prob(\bar A_{m})\prob(\bar A_{m-1}|\bar A_{m}).
\end{eqnarray*}
Assuming the BEC, $\prob(\bar A_{m-1}|\bar A_{m})=1$. Note that $\prob(\bar A_{m})=P_B^{(m)}$, where $P_B^{(m)}$ is the finite-length average block erasure rate $P_B$ at transmission $m$.
Remind that the expression for $P_B$ is given by (\ref{eq:app}).
 Therefore, we have for $\omega_m$
\begin{equation}
\label{eq:omega}
\omega_m=P_B^{(m-1)}-P_B^{(m)}.
\end{equation}
Note that (\ref{eq:omega}) is not valid for a more general type of transmission channel, where a subsequent
transmission may result in a more noisy version of the codeword (whereas for the BEC, each subsequent transmission can only
bring additional useful information).
However, (\ref{eq:omega}) could still be used as an approximation of $\omega_m$ in a more general case.

\begin{figure}[h]
\begin{picture}(0, 170)
\put(40,-10){\includegraphics[scale=0.5]{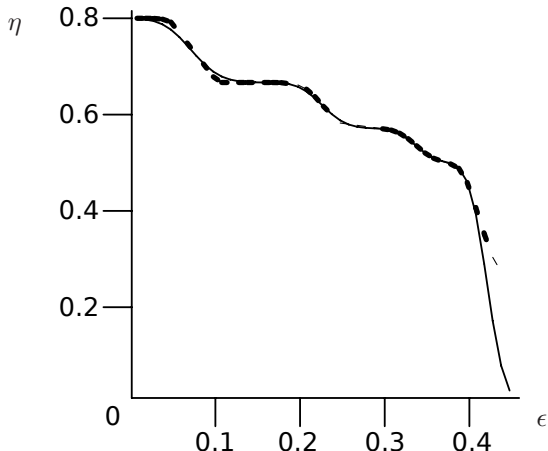}}
\put(20,150){$\eta$} \put(220,0){$\epsilon$}
\end{picture}
\caption{ \label{fig:bec}Average throughput $\eta$ versus equivalent channel erasure probability for $(x^2,x^5)$ LDPC codes of length $1024$. $M=5$. Dotted line - numerical results,  solid line - the analytical approximation.}
\end{figure}

By using Conjecture \ref{conj:1} to approximate $P_B^{(m)}$ in (\ref{eq:omega}), one gets an approximation of $\eta$ and $\tau$ for the IR-HARQ scheme. To support the use of Conjecture \ref{conj:1}, we present here a figure from \cite{AS09} that shows a good match of the approximation to numerical results.
In Figure \ref{fig:bec}, the average throughput of the IR-HARQ scheme with $M=5$, based on regular $(x^2,x^5)$ LDPC codes of length $1024$, is compared with its analytical approximation.

\section{Performance Optimization}
\label{sec:optimization}

Using the proposed puncturing model, we aim to optimize the performance of the IR-HARQ transmission scheme based on LDPC codes by
deciding which bits  should be sent at each transmission.
Note that, thanks to the concentration result for punctured LDPC codes, one has only to choose the mother LDPC code and the puncturing degree distributions for each transmission, without choosing a particular LDPC code and/or particular puncturing patterns.
The concentration of the punctured LDPC ensemble ensures that the performance of a particular punctured LDPC code, picked at random from the designed ensemble, will be close to the average performance of this ensemble with high probability. Thus our optimization problem is only to chose
how many bits on average should be sent in each transmission, rather than which exact bits.

The performance measure that we choose to optimize are the average throughput $\eta$ and the average delay $\tau$.
In previous sections, we have seen that, for finite-length schemes, $\eta$ has a staircase behavior, and thus it can be optimized point-wise, i.e., for some particular operating points on the $\epsilon$-axis, one optimizes $\eta$ to obtain the maximum possible throughput for those points.

We begin by assuming that the estimates of the erasure probabilities $\epsilon_1,\ldots, \epsilon_M$ are available at the transmitter.
We also fix the acceptable block erasure probability $P_B^{(M)}$ after the maximum number of transmissions\footnote{ In practice, $P_B^{(M)}$ is dictated by the supported application, i.e., image or voice transmission, video streaming, etc.
} $M$ and the feedback propagation delay $t$.

In the following section, we discuss the choice of other parameters that should be fixed before the optimization, namely: a) the maximum number of transmissions $M$, b)  the codelength $n$, c) a fixed or maximum transmission block size $K$  and d) the mother $(\lambda, \rho)$ LDPC code ensemble.
Then we investigate how to choose the puncturing degree distribution for each transmission $m$, $1\leq m < M$, which leads us to design a rate-adaptable punctured LDPC ensemble, based on the initial $(\lambda, \rho)$ ensemble and then adapted to transmission conditions.
Finally, we discuss how to obtain an estimate of erasure probabilities if they are not available at the transmitter.

\subsection{Choosing the Parameters $(\lambda,\rho)$, $n$, $M$ and $K$}
\label{sec:choice-param}
In this section, we discuss how one should go about
choosing the parameters $\lambda(x)$, $\rho(x)$, $n$, $M$ and $K$,
which in general depends on the anticipated IR-HARQ application.
The choice of degree distributions $\lambda(x)$ and $\rho(x)$ of the mother LDPC ensemble determines
the iterative decoding threshold $\epsilon^*$ and the code rate $R$ of the ensemble, and consequently,
an upper bound on the region of attainable throughputs versus transmission erasure probability.
See Fig.~\ref{fig:upbound} and Theorem \ref{thm:bounds}.

The upper bound on the region of attainable throughputs versus transmission erasure probability
achievable when $M \rightarrow n$ and $n$ is sufficiently large. Clearly, for practical schemes, i.e., for small values of $M$ and finite $n$ of order of several hundreds/thousands of bits, the average throughput is smaller.
However, if the degree distributions $\lambda(x)$ and $\rho (x)$ are chosen in such a way that $\epsilon^* > \max(\epsilon_1,\ldots,\epsilon_M)$  and the design rate $R$ is sufficiently large, they can be good initial choices for finite-length performance optimization.
Finally note that, if the desired block erasure probability $P_B^{(M)}$ is very low (e.g., $10^{-5}$ or lower,
depending on the code), this imposes additional constraints on the minimum distance of the code ensemble, and hence on the degree distributions $\lambda(x)$ and $\rho(x)$. Concerning these additional constraints, see, for example, \cite{Di-weight}.
The choice of the codelength $n$ depends on the desired value of $P_B^{(M)}$, which should be attainable for the given $\epsilon$ and the chosen $(\lambda,\rho)$-pair. This can be verified using the finite-length analysis from \cite{AMRU04}.

The maximum number of retransmissions $M$ should be chosen depending on a) the coherence time and b) the delay penalty.  The coherence time $T_C$ is the time during which the channel conditions are the same, and it depends on the transmission environment.
Note that in our model the instantaneous erasure probability $\epsilon_m$ is assumed to be constant during the $m$-th transmission.
Therefore, knowing $t_{1bit}$, we can transmit no more than $ \frac{T_C}{t_{1bit}}$ bits in one transmission.
From here we obtain that $M>\frac{t_{1bit} n}{T_C}$.
Since the delay penalty is proportional to the total time of feedback transmissions needed to transmit a packet of data,
to keep the delay penalty low one should choose $M$ so that the time of one single transmission, proportional to $\frac{n}{M}$, is large compared to the feedback propagation delay $t$.

In practice, the number $K$ of bits sent during one transmission is usually a constant, dictated by the transmission protocol. However, some applications may allow a variable length for the transmission block. To cover both cases, we define $K$ as the constant transmission block length in the first case and the maximum transmission block length in the second case.
Most often, $K$ is fixed and chosen to be $K = \frac{n}{M}$.

\subsection{Cost Function With a Feedback Penalty}
We next modify our optimization problem to address the case when the feedback transmission is not instantaneous but happens with
some delay $t$. This delay introduces the feedback penalty into the IR-HARQ transmission, which can be accounted for
in the average delay expression as explained by Remark~\ref{remark1}.

We start by defining a cost function, which needs to be optimized in order to increase the average throughput and to decrease the average delay.
From (\ref{eq:eta}) and (\ref{eq:tau2}), the average throughput and delay can be written as
$$\eta = \frac{R}{W_0}, \quad \quad \quad \tau = n W,$$
where
$$W_0 = \frac{1}{n} {\mathbb E}[ \#(\text{sent bits})| \text{ successful decoding, }n_{\text{\sc ACK}}=0]$$ and \\$W = \frac{1}{n} {\mathbb E}[ \#(\text{sent bits})| \text{ successful decoding, }n_{\text{\sc ACK}} \not = 0]$.
Thus,
$$W_0=\frac{\sum_{m=1}^M \omega_m \sum_{i, j} \lambda_i q_{ij} }{\sum_{m=1}^M \omega_m}$$
and $$ W = W_0 + \frac{\frac{n_{\text{\sc ACK}}}{n} \sum_{m=1}^M m \omega_m}{\sum_{m=1}^M \omega_m}.$$
Note that, having expressed $\eta$ and $\tau$ in terms of the same function $W_0$, one can see that the average throughput is inversely proportional to the average delay.
Moreover, if there is no feedback penalty, then $W=W_0$ and there is no tradeoff between optimizing the throughput and the delay: one achieves both goals by minimizing $W_0$. In the general case, when $W>W_0$, either $W_0$ or $W$ can serve as the cost function for the optimization problem. By choosing $W_0$, one ensures the optimum choice of $p_{im}$ coefficients to maximize the average throughput, and then $W$ is chosen to minimize the average delay.
Note that the solutions of two optimization problems, defined in terms of $W_0$ and $W$, are close to each other if the value of $n_{\text{\sc ACK}}$ is small compared to $n$.

From now on, we choose $W$ as the cost function for the optimization problem.
Using (\ref{eq:omega}) and (\ref{eq:pq}), $W$ can be rewritten in terms of the $p_{ij}$'s and $P_B^{(m)}$'s as:
\begin{align*}
W&=\frac{1}{1-P_B^{(M)}}\Bigl[( (1-P_B^{(M)})+\frac{n_{\text{\sc ACK}}}{n}\sum_{m=0}^{M-1}(P_B^{(m)}-P_B^{(M)})\\
&-\sum_{i} \lambda_i p_{i 1} + \sum_i \lambda_i \sum_{m=1}^{M-1} P_B^{(m)} (p_{im}-p_{i(m+1)}) \Bigr].
\end{align*}
Letting $\bar p_{j}= \sum_i \lambda_i p_{ij}$, we finally obtain
\begin{align}
\label{eq:W}
W&=\frac{1}{1-P_B^{(M)}} \Bigl (1-P_B^{(M)})+\frac{n_{\text{\sc ACK}}}{n}\sum_{m=0}^{M-1}(P_B^{(m)}-P_B^{(M)})\nonumber \\
&-\bar p_{1} + \sum_{m=1}^{M-1} P_B^{(m)} (\bar p_{m}-\bar p_{m+1}) \Bigr],
\end{align}
with
$P_B^{(m)}$ given by  (\ref{eq:app}) for puncturing degree distribution $p(x)=p_m(x)$ and for the average channel erasure probability $\epsilon=\delta_m$.
Note that, following (\ref{eq:eps-m}),
the average erasure probability $\delta_m$  through transmission $m$ is given by:
\begin{equation}
\delta_m =  \sum_i \lambda_i \frac{\sum_j (p_{i(j-1)}-p_{ij})\epsilon_j }{1-p_{im}}.
\end{equation}

\subsection{Optimization of Puncturing Degree Distributions}
Assuming the channel erasure probabilities $\epsilon_1,\ldots,\epsilon_M$ are known at the transmitter,
the optimization problem reduces to optimizing the puncturing degree distributions $p_m(x)=\sum_i p_{im} x^{i-1}$, $1 \leq m \leq M-1$, under the constraint of rate-compatibility, i.e.
\begin{multline*}
\text{argmin}_{p_{im}} W \quad
\text{for } \forall i \text{ and } \ 1\leq m \leq M-1,  \\ \text{given }1\geq p_{i1} \geq p_{i2} \geq \ldots \geq p_{i(M-1)}\geq p_{iM}=0.
\end{multline*}
In general, this is a non-linear optimization problem, given that $P_B^{(m)}$ depends on the parameters $\epsilon^*_m, \alpha_m$ and $\beta_m$, which themselves are dependent on the $p_{im}$'s.
We propose to use a gradient descent optimization algorithm to find a solution,
as described below.\\
{\tt START}\hfill\\
{\tt Initialization}\hfill
\begin{quotation}
For $m$ from $1 $ to $ M-1$, find initial puncturing fractions $\tilde p_{im}$'s by assuming that the iterative threshold $\epsilon^*_m$, given by (\ref{eq:de}), satisfies $\epsilon_m^* \geq \epsilon_m$.
Moreover, the $\tilde p_{im}$'s should satisfy one of the following conditions on $K$:
\end{quotation}
\begin{equation} \label{eq:init}
\sum_i^k (\tilde p_{i(m-1)}-\tilde p_{im}) = \frac{K}{n} \text{ or } \sum_i (\tilde p_{i(m-1)}-\tilde p_{im}) \leq \frac{K}{n}
\end{equation}
\begin{quotation}
for constant or variable transmission block size, respectively. \\
Choose the algorithm step size $\Delta_{max}$.
\end{quotation}
{\tt Main part}\hfill
\begin{quotation}
For $m$ from 1 to $M-1$, {\bf do} the following
iteration until the optimization process converges:
\end{quotation}
\begin{enumerate}
\item Using (\ref{eq:W}), compute $W$, given $p_{im}=\tilde p_{im}$, $\forall i$.
\item Find the $\Delta_{im}$'s that minimize
\begin{equation} \label{eq:deltaW} \Delta W = \sum_{i} \Delta_{im} \frac{\partial W}{\partial p_{im}} (\tilde p_{im})
\end{equation}
under the following constraints:
\begin{enumerate}
\item Maximum changes: $| \Delta_{im} |\leq \Delta_{max}$
\item Rate-compatibility: \[ 0 \leq   \tilde p_{im}+\Delta_{im}  \leq \tilde p_{i(m-1)}, \ \forall i\]
\item Number of bits sent per transmission:
\begin{align*}
 &\sum_i (\tilde p_{i(m-1)}-\tilde p_{im}-\Delta_{im})= \frac{K}{n}\\
 &\text{or} ~ \sum_i (\tilde p_{i(m-1)}-\tilde p_{im}-\Delta_{im}) \leq \frac{K}{n},
\end{align*}
for constant or variable block size.
\end{enumerate}
\item Set $\tilde p_{im} = \tilde p_{im}+\Delta_{im}$.
\end{enumerate}
\begin{quotation}
End of cycle over $m$.
\end{quotation}
{\tt Final part}\hfill
\begin{quotation}
\indent Set the puncturing fractions equal to $\tilde p_{im}$, $\forall i,m$.
\end{quotation}
{\tt END}\hfill\\[1mm]

Below are some details concerning the algorithm:
\begin{itemize}
\item {\it Initialization of $p_{im}$'s and choice of $\Delta_{max}$}:
The initial values of the puncturing fractions are proposed to be set as if the LDPC code were of infinite length.
This is an optimistic choice for the $\tilde p_{im}$'s, since a finite-length code will behave worse than an infinite-length one with the same parameters.
The fractions are found by linear programming: namely, one chooses puncturing fractions to maximize the code rate of the punctured ensemble, under the conditions of (\ref{eq:init}).
For more details on the optimization procedure, see, for instance, \cite{HKM04:rate-compatible}.
Note that, for small $m$ and high values of $\epsilon$, a solution may not exist. This means that the decoder will fail independently of the chosen puncturing fractions. In this case, any puncturing fractions can be chosen, assuming that they are rate-compatible with the optimized puncturing fractions for the later transmissions.
Such an initial choice for the puncturing fractions ensures good convergence for the gradient descent algorithm, since it already lies close to an optimal solution (see Conjecture \ref{conj:1} and Remark \ref{remark:1}).  Hence, the algorithm step size $\Delta_{max}$ should be chosen quite small, close to $\frac{1}{n}$.
\item {\it Minimization of (\ref{eq:deltaW})}: $\frac{\partial W}{\partial p_{im}}$ is given by
\begin{eqnarray}
\label{eq:wp}
\frac{\partial W}{\partial p_{im}} = \begin{cases}
(*),& m=1,\\
-c \lambda_i P_B^{(M-1)},&m=M,\\
(**),&1<m<M,\end{cases}
\end{eqnarray}
with
$(*)= -c \lambda_i\left(2- P_B^{(1)} \right)+c  \frac{\partial P_B^{(1)}}{\partial p_{i1}} \left( \bar p_1+\frac{n_{\text{\sc ACK}}}{n}\right)
$ and
$(**)=-c\lambda_i( P_B^{(m-1)}- P_B^{(m)})+c  \frac{\partial P_B^{(m)}}{\partial p_{im}}\left( \bar p_m+\frac{n_{\text{\sc ACK}}}{n}\right),
$ where $c=(1-P_B(M))^{-1}$ is a constant,
\begin{multline}
\frac{\partial P_B^{(m)}}{\partial p_{im}} = -\frac{\sqrt{n} \cdot exp\{\frac{n}{2}(\epsilon^*_m - \epsilon_m-\beta_m n^{-2/3})^2\}}{\sqrt{2 \pi} \alpha_m^2} \\ \left[ \alpha_m \left(\frac{\partial \epsilon^*_m}{\partial p_{im}} - n^{-2/3} \frac{\partial \beta_m}{\partial p_{im}}\right)
\right.\\\left.
- \frac{\partial \alpha_m}{\partial p_{im}}(\epsilon^*_m-\epsilon_m - n^{-2/3}\beta_m)\right],
\end{multline}
and $\epsilon^*_m$, $\alpha_m$ and $\beta_m$ are parameters of the LDPC ensemble, punctured corresponding to the puncturing polynomial $p_m(x)$. $\frac{\partial \alpha_m}{\partial p_{im}}$ and $\frac{\partial \beta_m}{\partial p_{im}}$ can be found by taking the derivative of (\ref{eq:alpha}) and (\ref{eq:beta}), and $\frac{\partial \epsilon^*_m}{\partial p_{im}}$ is obtained by implicit differentiation of the density evolution equation
\begin{equation}
\frac{\partial \epsilon^*_m}{\partial p_{im}} =  \frac{\lambda_i y_m^{i-1}( x_m-\lambda(x_m))  }{\bar \lambda_p(x_m)^2}.
\end{equation}
\end{itemize}

\begin{remark}
Note that the optimization problem based on $W_0$ instead of $W$ is defined in exactly the same way, except that the terms $n_{\text{\sc ACK}}/n$ will in (\ref{eq:wp}) will be zero.
\end{remark}

\subsection{An Example of Optimization}
\label{sec:example-irreg}
Now we consider a particular example of the optimization of an LDPC ensemble for a particular value of the channel average erasure probability $\epsilon_{target}$.
The initial parameters are: $n=2000$, $M=5$, $P_B(M)=0.01$ and $K=n/M=400$, where the transmission block size $K$ is constant.
Denote by $\epsilon_{max}$ the maximum erasure probability that can be tolerated by the LDPC ensemble.
We will choose $\epsilon_{target} = 0.35$ and $\epsilon_{max}=0.55$
and optimize the throughput at $\epsilon_{target} = 0.35$ under the constraint that the iterative decoding threshold $\epsilon^* \geq \epsilon_{max}$.

The following degree distributions were chosen:
$\lambda(x) =0.220813x+  0.353686x^3+0.425502x^{12}$ and
$\rho(x) =0.390753x^4+0.361589x^5+0.247658x^9$.
This gives rise to an LDPC ensemble with rate $R=0.37$, $\epsilon^*=0.608$ (from (\ref{eq:de})) and $P_B(M=5, n=2000, \epsilon_{max})\approx0.009$ (from (\ref{eq:app})).
The optimized puncturing degree distributions at the initialization stage are
\begin{align*}
&\tilde p_4(x)=0.6x,\\
&\tilde p_3(x)=0.60264x+0.123057x^3+0.474303x^{12},\\
&\tilde p_2(x)= 0.735093x+0.415371x^3+0.649536x^{12},\\
&\tilde p_1(x)=0.867547x+0.707686x^3+0824768x^{12}.
\end{align*}
We find that $P_B^{(m)}=1$ for $m\leq 2$, i.e. after the first two transmissions a decoder will fail 
because of an insufficient number of transmitted bits, no matter what puncturing degree distributions are used.
$\tilde p_3(x)$ and $\tilde p_4(x)$, however, are the best choices for the given initial parameters.
Therefore, one needs to do at least 3 transmissions before starting to decode. Knowing this, we can send the first three coded packets one after another, without  waiting for the feedback.

\begin{figure}[h]
\begin{picture}(0, 170)
\put(0,0){\includegraphics[scale=0.85]{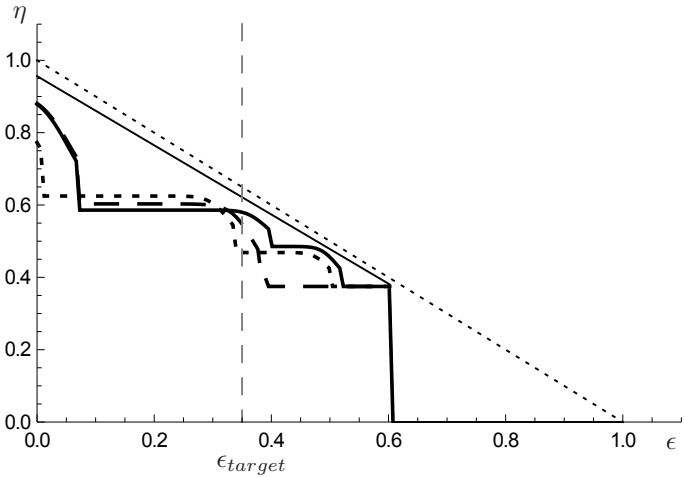}}
\put(3,165){$\eta$} \put(250,2){$\epsilon$}
\put(80,-5){$\epsilon_{target}$}
\end{picture}
\caption{ \label{fig:ex-irreg} Impact of the choice of $p_m(x)$'s on the average throughput. The thin dotted line represents the theoretical limit, the thin full line - the upper bound given the mother LDPC code, the thick dotted line - no optimization, the thick dashed line - infinite-length optimization, the thick full line - finite-length optimization.}
\end{figure}

For the initial-stage $p(x)$'s, the cost function $W = 0.677$. After the finite-length optimization, we obtain $W = 0.646$ with the following new distributions $\tilde p_3(x)$ and $\tilde p_4(x)$:
\begin{eqnarray*}
\tilde p_4(x)&=&0.1351x+0.4649x^{12},  \\
\tilde p_3(x)&=&0.7351x+0.4649x^{12}.
\end{eqnarray*}
The average throughput, obtained using the described optimization procedure, is shown by the thick full line in Figure \ref{fig:ex-irreg}.
The throughput with puncturing degree distributions obtained at the initialization stage is shown by the thick dashed line.
Also, the thick dotted line represents the average throughput, obtained without any optimization by equally partitioning the bits of each degree $i$ between transmissions.
As we can see, the throughput at $\epsilon_{target} = 0.35$ has indeed been improved.

This example illustrates the interesting point that, in order to obtain a higher average throughput for some $\epsilon_{target}$, one should not blindly send the bits with higher degrees first, trying to get the iterative decoder converge faster (which would seem intuitive), but instead find the optimal puncturing degree distributions for the given $\epsilon_{target}$. The reason is the following: if one of the first transmissions, carrying a large number of high degree bits, is unsuccessful, it will cause a large fraction of those bits to be erased, and many more transmissions will be needed in order to accumulate a sufficient number of unerased bits with lower degrees to make the decoder converge.

Note that one can define an optimization problem for more than one target erasure probability, thus optimizing the throughput curve pointwise.
Also note that the parameter $K$ operates as a regulator of the number of transmissions. If the number of sent bits per packet were unbounded, there would be at most 2 transmissions -- for the first transmission,
the optimizer would decide to send as many bits as needed to ensure the target $P_B$ at a given $\epsilon_{target}$, and, if the first transmission were unsuccessful, it would allocate the rest of the bits to transmission 2.

\subsection{The Regular Code Case}
\label{sec:example-reg}
In the case of regular LDPC codes, the scaling and shift parameters do not depend on the puncturing fraction $p_m$. Indeed, based on Example \ref{example:reg}, it is easy to see that, for punctured regular codes,
\begin{equation}
P_B^{(m)}= Q\left( \frac{\sqrt{n}(\epsilon^*-\epsilon(1-p_m)-\beta n^{-2/3})}{\alpha}\right),
\end{equation}
where $\epsilon^*$, $\alpha$ and $\beta$ are parameters of the initial unpunctured regular ensemble.
Since $Q(x)$ is an increasing function of its argument, $P_B^{(m)}$ is a monotone increasing function in $p_m$ and the cost function $W$ is minimized by the smallest possible values of $p_m$, $1\leq m <M$.

\subsection{Estimating the $\epsilon_m$'s at the Transmitter}
In general, the channel erasure probabilities $\epsilon_1,\ldots,\epsilon_M$ are not known at the transmitter and must be estimated before performing the optimization of the puncturing degree distributions.
The quality of estimation depends on the knowledge of the transmission channel statistics (mean, variance, probability distribution) and on the amount of feedback obtained at the transmitter (1 bit representing an ACK/NACK, the previous channel erasure probability,...).

A wealth of literature is available on channel estimation.
As examples, we list below a few possible approaches to channel estimation.

\begin{itemize}
\item {\it Known mean}: Let the mean $\bar \epsilon$ of the channel erasure probability be known at the transmitter. Then the puncturing degree distribution can be optimized as discussed above, assuming $\epsilon_m = \bar \epsilon$, $m=1,\ldots,M-1$.
\item {\it Known mean and previous erasure probabilities}:
Let the mean $\bar \epsilon$ of the channel erasure probability be known and assume the receiver sends to the transmitter the erasure probabilities  $\epsilon_1,\ldots,\epsilon_{m-1}$ of the previous transmissions.
In this case one can optimize the puncturing degree distributions in real time, i.e., just prior to transmission.
At transmission $1$, the transmitter sends the fraction of coded bits, optimized for $\epsilon_1=\bar \epsilon$, since it does not have any feedback information. At transmission $m>1$, however, the estimated erasure probability becomes
$$\epsilon_m = m \bar \epsilon - \sum_{i=1}^{m-1} \epsilon_i.$$
\item {\it Known probability distribution and 1-bit feedback}:
Assume that the probability density function $p(\epsilon)$ is known and it has support $[\epsilon_{min},\epsilon_{max}]$.
Then, for each transmission $m$, we can estimate
\begin{multline*}\epsilon_m = \text{argmax}_{\epsilon \in [\epsilon_{min},\epsilon_{max}]} \\ \Pr(\epsilon = \epsilon_m|
\text{\sc ACK/NACK}_1, \ldots, \text{\sc ACK/NACK}_{m-1}).\end{multline*}
Also note that, to ensure good performance, one should choose $\lambda(x)$ and $\rho(x)$ in such a way that $\epsilon^*\geq \epsilon_{max}$.
\end{itemize}

\section{Rateless Incremental Redundancy Protocols}
\label{sec:new}

\subsection{Rateless Protocols Using Repetition}
As can be seen in Figure \ref{fig:upbound}, the IR-HARQ protocols based on punctured codes
achieve a high throughput only over a limited region of channel erasure rates. When they are based on
iterative decoding and a mother LDPC code with threshold $\epsilon^*$, this region extends from $0$ to $\epsilon^*$.
Naturally, to cover a larger region, one can choose a mother LDPC code with $\tilde \epsilon^*> \epsilon^*$. However,
such a code may have a lower rate $\tilde R < R$, and moreover $\tilde \eta(0)=\frac{\tilde R}{1-\tilde \epsilon^*}$ may be lower than $\eta (0)=\frac{R}{1-\epsilon^*}$, resulting in a lower throughput in the region $\epsilon<\epsilon^*$ (see Figure~\ref{fig:upbound}). Compare, for example, the rate $1/2$ regular $(x^2,x^5)$ code with $\epsilon^*=0.4293$ and $\eta(0)=0.876$ to the
rate $2/5$ regular $(x^2,x^4)$ code with $\epsilon^*=0.5176$ and $\eta(0)=0.829$.

To extend the region of high throughput for a given mother code, we propose to augment the HARQ protocol as follows.
If, after the transmission of all the bits in the codeword, decoding still fails, we further increment redundancy
simply by repeating the same codeword, using the same $q_{im}$.
Hence, each coded bit might be transmitted twice through channels with erasure probabilities $\epsilon^{(1)}$ and $\epsilon^{(2)}$. At the receiver, both received values of a bit are combined together. So, after two transmissions, the bit is erased with probability $\epsilon^{(1)} \epsilon^{(2)}$. One can continue transmitting in this manner, making the scheme essentially rateless.

The proposed protocol is called the incremental redundancy protocol with repetition, and we denote it by IR-Rep-HARQ. Although repetition is in general not optimal, note that it takes place only when the channel conditions are bad ($\epsilon > \epsilon^*$), when it actually is a good strategy to follow. Note that in the repetition stage, we can either retransmit the same blocks as in the first stage, or determine new blocks, according to the optimized fractions $\{q_{im}\}$. This translates to generating new $\theta$ values in the protocol of Section \ref{sec:mult-transm}.
We next find expressions for the average throughput and the average delay for these two cases.

\begin{enumerate}
\item {\it Repetitions of the same blocks}\\
Assume the IR-Rep-HARQ protocol with repetitions of the same blocks during the second transmission of the codeword.
Denote the channel erasure probabilities by $\epsilon_1^{(1)},\ldots,\epsilon_M^{(1)}$ for the first transmission and by $\epsilon_1^{(2)},\ldots,\epsilon_M^{(2)}$ for the second transmission.
Then, similar to  (\ref{eq:eta}) and (\ref{eq:tau}), the average throughput $\eta_{\text{IR-Rep}}$ and the average delay $\tau_{\text{IR-Rep}}$ are given by
\begin{eqnarray}
\label{eq:eta-rep}
\eta_{\text{IR-Rep}} &=& \frac{R\displaystyle{\sum_{r=1}^2 \sum_{m=1}^{M}\omega_m^{(r)}}}{\displaystyle{\sum_{r=1}^2 \sum_{m=1}^{M} \omega_m^{(r)}
\Bigl(\sum_{j=1}^m \bar q_j\Bigr)}},
\\
\label{eq:tau-rep}
\tau_{\text{IR-Rep}} &=& \frac{n \displaystyle{\sum_{r=1}^2 \sum_{m=1}^{M} \omega_m^{(r)}
\Bigl(\sum_{j=1}^m \bar q_j\Bigr)}}{\displaystyle{\sum_{r=1}^2 \sum_{m=1}^{M}\omega_m^{(r)}}},
\end{eqnarray}
where $\bar q_j = \sum_i \lambda_i q_{ij}$ and $\omega_m^{(r)} = P_B(\delta_{m-1}^{(r)})-P_B(\delta_{m}^{(r)})$, with
\begin{equation} \label{eq:eps-m-r}
\delta_m^{(r)}= \begin{cases}
\frac{\sum_{j=1}^m \bar q_{j} \epsilon_j^{(1)} }{\sum_{k=1}^m \bar q_{k}},& r=1;\\
\frac{\sum_{j=1}^m \bar q_{j} \epsilon_j^{(1)}\epsilon_j^{(2)} }{\sum_{k=1}^m \bar q_{k}}+\frac{\sum_{j=m+1}^M \bar q_{j} \epsilon_j^{(1)}}{{\sum_{k=m+1}^M \bar q_{k}}}
,& r=2.
\end{cases}
\end{equation}

Or, equivalently,
\begin{eqnarray}
\label{eq:eta-rep2}
\eta_{\text{IR-Rep}} =
 \eta_{r=1} (1-P_B(\delta_M^{(1)}))+ \eta_{r=2}P_B(\delta_M^{(1)}),
\end{eqnarray}
where $\eta_{r=1} = \eta$ is given by (\ref{eq:eta}) and
$$\eta_{r=2}=\frac{R\displaystyle{\sum_{m=1}^{M}\omega_m^{(2)}}}{\displaystyle{\sum_{m=1}^{M} \omega_m^{(2)}
\Bigl(\sum_{j=1}^m \bar q_j\Bigr)}}.$$
Similarly,
\begin{eqnarray}
\label{eq:tau-rep2}
\tau_{\text{IR-Rep}} &=& \tau (1-P_B(\delta_M^{(1)}))+\nonumber \\&& \frac{2n \displaystyle{ \sum_{m=1}^{M} \omega_m^{(2)}
\Bigl(\sum_{j=1}^m \bar q_j\Bigr)}}{\displaystyle{\sum_{m=1}^{M}\omega_m^{(2)}}} P_B(\delta_M^{(1)}),
\end{eqnarray}
where $\tau$ is given by (\ref{eq:tau}).
\item {\it Repetition with different blocks}\\
Assume the IR-Rep-HARQ protocol such that the repetition of the block $m$ is chosen at random from the available, non-repeated bits, according to the fractions $\{q_{im}\}$.
Then the expressions for $\eta_{\text{IR-Rep}}$ and $\tau_{\text{IR-Rep}}$ are the same as in the previous case (see (\ref{eq:eta-rep}) and (\ref{eq:tau-rep})), except that the equivalent average erasure probability $\delta_m^{(r)}$ is computed as
\begin{equation} \label{eq:eps-m-r-2}
\delta_m^{(r)}= \begin{cases}
\frac{\sum_{j=1}^m \bar q_{j} \epsilon_j^{(1)} }{\sum_{k=1}^m \bar q_{k}},& r=1;\\
A
,& r=2,
\end{cases}
\end{equation}
with $$A=\sum_{j=1}^m \bar q_{j}\epsilon_j^{(2)} \large(\sum_{k=1}^M \bar q_k \epsilon_j^{(1)} \large) +\sum_{j=m+1}^M (1-\bar q_{j} )\large( \sum_{k=1}^M \bar q_k \epsilon_j^{(1)}\large).$$
\end{enumerate}

We now develop bounds on $\eta_{\text{IR-Rep}}$ and $\tau_{\text{IR-Rep}}$ for these two IR-Rep-HARQ schemes.
{\theorem The average throughput $\eta_{\text{IR-Rep}}$ for the IR-Rep-HARQ schemes
with the same or different repeated blocks
is bounded by
\begin{equation} \label{eq:UB-eta-rep}
\eta_{\text{IR-Rep}} \leq \begin{cases}
\frac{R(1-\epsilon)}{1-\epsilon^*_{(n)}},& \epsilon \leq \epsilon^*_{(n)};\\
\frac{R}{1+\frac{\epsilon-\epsilon^*_{(n)}}{\epsilon-\epsilon^2}},& \epsilon^*_{(n)}<\epsilon \leq \sqrt{\epsilon^*_{(n)}};\\
0,& \sqrt{\epsilon^*_{(n)}}< \epsilon \leq 1.
\end{cases}
\end{equation}
}
\begin{proof}
Assume $M=n$.
For $\epsilon \leq \epsilon^*_{(n)}$ (transmission without repetition), the expression for $\eta$ is already given by Theorem~\ref{thm:bounds}.
Consider now $\epsilon > \epsilon^*_{(n)}$ (transmission with repetition).
Let some fraction $\gamma$ of bits be sent twice, $0 \leq \gamma \leq 1$.
Note that sending bits twice is equivalent to sending them over a BEC with erasure probability $\epsilon^2$.
Hence the equivalent erasure probability $p$ is given by
$p=(1-\gamma)\epsilon+\gamma \epsilon^2.$
There are two possible cases to consider. If $p> \epsilon^*_{(n)}$, the average throughput $\eta_{\text{IR-Rep}}$ of the IR-Rep-HARQ scheme is $0$. If $p\leq \epsilon^*_{(n)}$, the average throughput is strictly positive and can be expressed as
$$\eta_{\text{IR-Rep}} = \frac{Rn}{2\gamma n + (1-\gamma)n} = \frac{R}{1+\gamma},$$
where $2\gamma n + (1-\gamma)n$ is the total number of sent bits and $Rn$ is the number of information bits.

Now we find the values of $\epsilon$ for which the throughput is positive. From the condition $p\leq \epsilon^*_{(n)}$ it follows that
$$\frac{\epsilon-\epsilon^*_{(n)}}{\epsilon-\epsilon^2} \leq \gamma \leq 1.$$
Thus $\eta_{\text{IR-Rep}}>0$ when
$\epsilon \leq \sqrt{\epsilon^*_{(n)}}$.
Moreover, we can upper bound $\eta_{\text{IR-Rep}}$ in the interval $\epsilon^*_{(n)}<\epsilon \leq \sqrt{\epsilon^*_{(n)}}$ by
$$\eta_{\text{IR-Rep}} = \frac{R}{1+\gamma} \leq \frac{R}{1+\frac{\epsilon-\epsilon^*_{(n)}}{\epsilon-\epsilon^2}}.$$
\end{proof}

The following lower bound on $\tau_{\text{IR-Rep}}$ can be derived using a similar approach.
{\theorem The average delay $\tau_{\text{IR-Rep}}$ for IR-Rep-HARQ schemes with the same or different repeated blocks is bounded by
\begin{equation} \label{eq:LB-tau-rep}
\tau_{\text{IR-Rep}} \geq \begin{cases}
\frac{n(1-\epsilon^*_{(n)})}{1-\epsilon},& \epsilon \leq \epsilon^*_{(n)};\\
n\left(1+\frac{\epsilon-\epsilon^*_{(n)}}{\epsilon-\epsilon^2}\right),& \epsilon^*_{(n)}<\epsilon \leq \sqrt{\epsilon^*_{(n)}};\\
\infty,& \sqrt{\epsilon^*_{(n)}}< \epsilon \leq 1.
\end{cases}
\end{equation}
}

\begin{figure}[t]
\begin{picture}(0, 170)
\put(20,0){\includegraphics[scale=0.4]{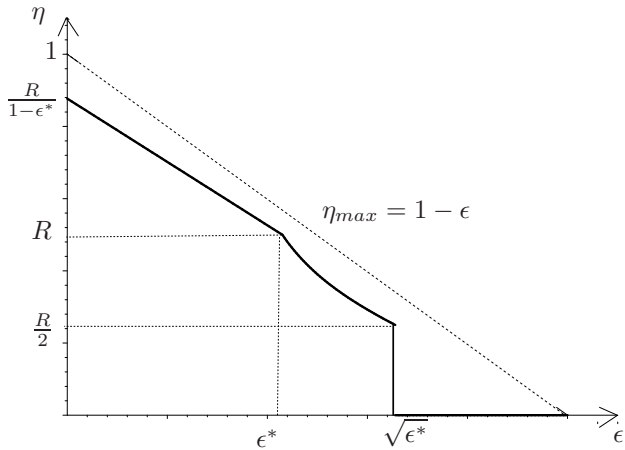}}
\put(10,155){$\eta$} \put(230,-5){$\epsilon$} \put(95,-7){$\epsilon^*$}
\put(15,140){$1$} \put(350,-3){$1$} \put(120,80){$\eta_{max}=1-\epsilon$}
\put(10,72){$R$} \put(10,35){$\frac{R}{2}$} \put(0,122){$\frac{R}{1-\epsilon^*}$} \put(143,-5){$\sqrt{\epsilon^*}$}
\end{picture}
\caption{ \label{fig:etarep} Upper bound on the throughput for an IR-Rep-HARQ scheme based on $(x^2,x^5)$ LDPC codes (black curve). The straight line above is the maximum attainable throughput.}
\end{figure}

As an example, the upper bound on the throughput for the scheme based on regular $(x^2,x^5)$ LDPC codes is shown in Fig.\ref{fig:etarep}. For simplicity, we assume a large codelength $n$ and $\epsilon^*_{(n)}\approx \epsilon^*$.
We see that, in the region of erasure probabilities from $\epsilon^*\approx 0.43$ to $\sqrt{\epsilon^*} \approx 0.63$, repetition of the same codeword results in an almost linear upper bound on throughput.

Extending the above results to IR-Rep-HARQ schemes with a larger number of repetitions is straightforward.
We state this extension without proof in the following corollary.
{\corollary
Consider an IR-Rep-HARQ scheme, based on LDPC codes, with $L$ repetitions.
Denote $(\epsilon^*_{(n)})^{1/r}$ by $\varepsilon(r)$.
Then the following bounds hold:
\begin{eqnarray*}
\eta_{\text{IR-Rep}} \leq \begin{cases}
R\frac{1-\epsilon}{1-\epsilon^*_{(n)}},&    \epsilon \leq \epsilon^*_{(n)} \text{ for } r=1;\\
R\left(1+\frac{\epsilon^{r-1}-\epsilon^*_{(n)}}{\epsilon^{r-1}-\epsilon^r}\right)^{-1},&
\varepsilon(r-1) < \epsilon \leq \varepsilon(r) \\&\text{ for } r=2,\ldots, L;\\
0,& \epsilon> \varepsilon(L);
\end{cases}\\
\tau_{\text{IR-Rep}} \geq \begin{cases}
n\frac{1-\epsilon^*_{(n)}}{1-\epsilon},&      \epsilon \leq \epsilon^*_{(n)} \text{ for } r=1;\\
n \left(1+\frac{\epsilon^{r-1}-\epsilon^*_{(n)}}{\epsilon^{r-1}-\epsilon^r}\right),&   \varepsilon(r-1) < \epsilon \leq \varepsilon(r)\\&\text{ for } r=2,\ldots, L;\\
\infty,& \epsilon> \varepsilon(L).
\end{cases}
\end{eqnarray*}
}

\subsection{Comparison with LT Codes}

It is natural to compare the performance of IR-HARQ schemes based on punctured LDPC codes with those based on other rateless codes.
We consider LT codes as an example.
Since an IR-HARQ-LT scheme does not have a maximum number of transmissions $M$, we assume that $M=n$ for the IR-HARQ-LDPC schemes, which leads us naturally to comparing the upper bounds on throughput of the two schemes.

Assume there are $K$ information bits to transmit. From Sec.~5 of \cite{Sh06}, the upper bound on throughput of the IR-HARQ-LT schemes under belief propagation decoding is given by $$\eta_{\text{FC-HARQ}} \leq \frac{1-\epsilon}{1+\frac{\log^2 K}{\sqrt{K}}}.$$
For the IR-HARQ-LDPC schemes, we must choose a code rate and a code ensemble.
As examples, we take two code ensembles already considered in the paper: regular $(x^2,x^5)$ LDPC codes of rate $1/2$ and irregular LDPC codes of rate $0.37$, optimized in Section \ref{sec:example-irreg}.

\begin{figure}[hbt]
\begin{picture}(0, 150)
\put(20,0){\includegraphics[scale=0.4]{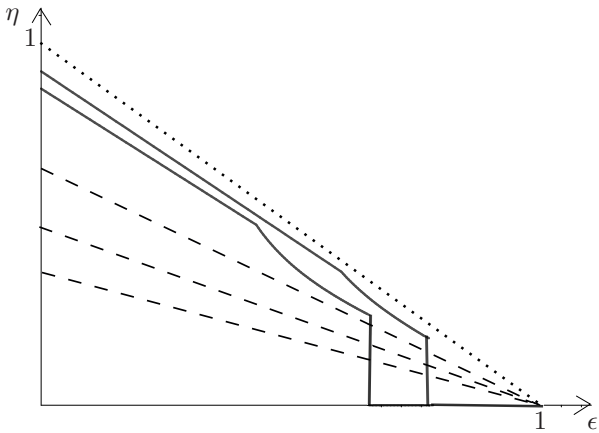}}
\put(10,150){$\eta$} \put(230,-5){$\epsilon$}
\put(17,140){$1$} \put(210,-5){$1$}
\end{picture}
\caption{ \label{fig:FC} Comparison of upper bounds on the throughput of IR-Rep-HARQ-LDPC schemes with one repetition, based on regular codes (lower solid curves) and on irregular codes (upper solid curves), and of the IR-HARQ-LT scheme  (dashed curves) with $K=500;5000;50000$. For IR-Rep-HARQ-LDPC, bounds virtually coincide for all $K$. For IR-HARQ-LT, the lowest curve corresponds to the smallest value of $K$. The dotted line corresponds to the maximum theoretical throughput.}
\end{figure}

Figure \ref{fig:FC} presents a comparison of two IR-Rep-HARQ schemes and of one IR-HARQ-LT scheme for different values of $K$.
Note that in the region of small values of $\epsilon$, the IR-HARQ-LDPC schemes  have better maximum throughputs than the IR-HARQ-LT schemes. Moreover, the throughput of the IR-HARQ-LDPC schemes can be improved using repetition for $\epsilon>\epsilon^*$.
Finally, for very poor channels ($\epsilon \approx 1$), the IR-HARQ-LT schemes have a better throughput than the double-repetition
IR-HARQ-LDPC schemes.

\section{Discussion and Future Work}
\label{sec:discussion}
We considered IR-HARQ schemes based on finite-length punctured LDPC codes, where the transmission was assumed to take place over the time-varying binary erasure channel, with the goal to characterize and optimize the
throughput and the delay obtained by using different puncturing degree distributions. 
Our goal was achieved by following two approaches: 1) approximating the block erasure performance of finite-length punctured LDPC codes used in computing the throughput and delay and 2) computing an upper (lower) bound on the average throughput (delay).
We also proposed an optimization algorithm for the puncturing degree distribution to improve the performance of the IR-HARQ protocol based on LDPC codes with finite number of transmissions. We introduced a transmission protocol, called Incremental Redundancy HARQ with Repetition (IR-Rep-HARQ), which extends the region of channel erasure rates over which good performance can be obtained with the IR-HARQ scheme.

There are three main {\sl contributions} of this paper:
\begin{enumerate}
\item We have defined a cost optimization function that minimizes the delay or maximizes the throughput in the case of small feedback overhead ($n_{\text{\sc ACK}}$). Optimizing this function (if the optimum exists) comes very close to achieving the best possible tradeoff between throughput and delay. The proposed cost function can be optimized point-wise, i.e., for a set of target channel erasure probabilities.
\item We have shown that, from the point of view of performance optimization, there is an important difference between using regular and irregular LDPC codes in IR-HARQ schemes. The cost optimization function is monotone for regular LDPC codes, and the puncturing degree distribution for each of transmissions is simple to calculate.
For irregular LDPC codes, the cost optimization function is not monotone, and the puncturing degree distributions must be carefully optimized in order to obtain the best throughput or delay.
\item We have deminstrated that each repetition of a complete IR-HARQ round upon a failure to decode improves the throughput and extends the region of channel erasure probabilities over which good performance can be obtained beyond the iterative threshold $\epsilon^*$ of the mother LDPC code. Hence an IR-HARQ scheme based on punctured sparse-graph codes can be made rateless with a high throughput.
In particular, an IR-HARQ scheme with repetitions based on punctured LDPC codes outperforms an HARQ scheme based on LT codes over a large region of channel erasure probabilities.
\end{enumerate}

It is important to note the following:
\begin{itemize}
\item {\sl All protocol stack:}
Our approximation of $P_B$ can also be combined with the reasoning of \cite{ciblat-packetHARQ} to obtain a more accurate expression of the failure probability, which would take into account failure events at all layers of the protocol stack.
\item {\sl Other types of channels:}  In principle, our results can be extended to other binary-input symmetric memoryless channels, thus modeling transmissions at the physical layer. In this case the estimation of parameters is more involved, but still feasible. As an alternative, they can be estimated numerically before the optimization algorithm is initiated.
\item {\sl Universality of the optimization algorithm:} The optimization algorithm is quite general and can be easily adapted to other scenarios, e.g., when the packet size varies or when feedback is only sent periodically rather than after each transmission (see time duplex division schemes in \cite{5062100}).
\item {\sl Tightness of the approximation:}  Note that the average throughput and the average delay, obtained for given $p_m(x)$, $1 \leq m \leq M$, is tight, owing to the tightness of the $P_B^{(m)}$ approximation. Of course, this approximation is valid only for the so called waterfall region of the performance curve. However, it is precisely this region that is of interest for practical HARQ schemes because of the very
    nature of the protocol.

\end{itemize}

The following extensions would be of interest:
\begin{itemize}
\item {\sl Accuracy of channel state prediction:} It would be of interest to consider various levels of channel state information (CSI) and to obtain the IR-HARQ protocol performance in each case.
\item {\sl Using other punctured codes:} The optimization is not limited to LDPC codes. It can be extended to other code ensembles for which ther exists a finite-length performance approximation, such as turbo-like codes, which could give insight into the design of code ensembles that perform well in particular retransmission protocols.
\item {\sl Finding the expressions for throughput and delay at each network layer:} This work can serve as a basis for optimizing network parameters at any network layer, once the expression of the failure probability, based on results from \cite{ciblat-packetHARQ}, is available.
\end{itemize}

\section*{Acknowledgment}
A significant part of this work was done in 2008 at Ecole Polytechnique F\'{e}d\'{e}rale de Lausanne (EPFL).
The authors would like to thank their EPFL hosts, professors A.~Shokrollahi and R.~Urbanke for their support, and also anonymous reviewers for helpful remarks and comments.

\bibliographystyle{IEEEtran}
\bibliography{punctured}

\end{document}